\documentclass{elsarticle}

\usepackage{lineno,hyperref}
\usepackage{times}  
\usepackage{helvet}  
\usepackage{courier}  
\usepackage{url}  
\usepackage{graphicx}  
\usepackage{amsmath}
\usepackage{amsfonts}
\usepackage{algpseudocode}
\usepackage{amsmath}
\usepackage{array}
\usepackage{graphics}
\usepackage{epsfig}
\usepackage{subfigure}
\usepackage{multirow}
\usepackage[misc]{ifsym} 

\usepackage{color}
\usepackage{floatrow}  
\newfloatcommand{capbtabbox}{table}[][\FBwidth]  

\newcommand{\tabincell}[2]{\begin{tabular}{@{}#1@{}}#2\end{tabular}}
\newtheorem{myDef}{\textit{Definition}}
\newtheorem{myPro}{\textit{PROBLEM}}

\journal{Big Data Research}

\begin{document}

\begin{frontmatter}

\title{SMR: Medical Knowledge Graph Embedding for Safe Medicine Recommendation}

\author[sh]{Fan Gong}
\author[seu,lab]{Meng Wang(\Letter)}
\author[tj]{Haofen Wang}
\author[uq]{Sen Wang}			
\author[xjtu]{Mengyue Liu}

\address[sh]{Shanghai Shuguang Hospital Affiliated to Shanghai University of Traditional Chinese Medicine, Pu’an Road, Shanghai, China}
\address[seu]{School of Computer Science and Engineering, Southeast University, Nanjing, China}
\address[lab]{Key Laboratory of Computer Network and Information Integration (Southeast University), Ministry of Education, Nanjing, China}
\address[tj]{College of Design and Innovation, Tongji University, Shanghai, China}
\address[uq]{The University of Queensland, Brisbane, Australia}
\address[xjtu]{School of Electronic and Information Engineering, Xi'an Jiaotong University, Xi’an, China}

\begin{abstract}
Most of the existing medicine recommendation systems that are mainly based on electronic medical records (EMRs) are significantly assisting doctors to make better clinical decisions benefiting both patients and caregivers. Even though the growth of EMRs is at a lighting fast speed in the era of big data, content limitations in EMRs restrain the existed recommendation systems to reflect relevant medical facts, such as drug-drug interactions. Many medical knowledge graphs that contain drug-related information, such as DrugBank, may give hope for the recommendation systems. However, the direct use of these knowledge graphs in systems suffers from robustness caused by the incompleteness of the graphs. To address these challenges, we stand on recent advances in graph embedding learning techniques and propose a novel framework, called Safe Medicine Recommendation (SMR), in this paper. Specifically, SMR first constructs a high-quality heterogeneous graph by bridging EMRs (MIMIC-III) and medical knowledge graphs (ICD-9 ontology and DrugBank). Then, SMR jointly embeds diseases, medicines, patients, and their corresponding relations into a shared lower dimensional space. Finally, SMR uses the embeddings to decompose the medicine recommendation into a link prediction process while considering the patient's diagnoses and adverse drug reactions. Extensive experiments on real datasets are conducted to evaluate the effectiveness of proposed framework.
\end{abstract}


\end{frontmatter}

\section{Introduction}
Over the last few years, medicine recommendation systems have been developed to assist doctors in making accurate medicine prescriptions. On the one hand, many researchers \cite{chen2016physician,almirall2012designing} adopt rule-based protocols that are defined by the clinical guidelines and the experienced doctors. Constructing, curating, and maintaining these protocols are time-consuming and labor-intensive. Rule-based protocols might be effective for a general medicine recommendation for a specific diagnosis, but give little help to tailored recommendations for complicated patients. On the other hand, supervised learning algorithms and variations, such as Multi-Instance Multi-label (MIML) learning \cite{zhang2017leap}, have been proposed to recommend medicines for patients. Both input features and ground-truth information that are extracted from massive EMRs are trained to obtain a predictive model that outputs multiple labels of the new testing data as medicine recommendations. It is a fact that therapies and treatments in clinical practices are rapidly updated. Unfortunately, supervised learning methods cannot deal with those medicines that are not included in the training phase. Incomplete training data set will be a detriment to the recommendation system performance. 

It is reported in \cite{panagioti2015multimorbidity,elderlypatients} that patients with two or more diseases, acute or chronic, often take five or more different medicines simultaneously and have immense health risks. Studies \cite{adverse,35errors} have shown that 3-5\% of all in-hospital misused prescriptions blame to ignorances of adverse drug reactions, which is difficult to prohibit even for the highly trained and experienced clinicians. With the assistance from the conventional medicine recommendation systems, clinicians still need to cautiously rule out those recommendations that have potential adverse effects caused by drug-drug interactions. Most of the existing works have largely ignored the exploit of medical facts in medicines, such as drug-drug interactions, which is crucial in medicine recommendation system. One possible reason might because there is little medical expert knowledge in EMRs. Content limitations in EMRs constrain the systems to barely associate accurate medical facts with the recommended prescriptions, which makes the final recommendation less trustworthy for the complicated patients.

With the increasing emergence of knowledge graphs, many world-leading researchers have successfully extracted information from huge volumes of medical databases to build up giant heterogeneous graphs that reflect medical facts of medicines and diseases. For instance, DrugBank \cite{drugbank} is a rich source of medicine information. It contains extensive entities (drugs, drug targets, chemistry, etc.) and relationships (enzymatic pathways, drug-drug interactions, etc.). ICD-9 ontology \cite{schriml2011disease} represents a knowledge base of human diseases and can be used to classify diagnoses of patients. Harnessing well-built medical knowledge graphs in EMRs-based medicine recommendation system might enable the reinvented system to provide appropriate prescriptions for special patients, as well as alerts of possible side effects and serious drug-drug interactions (DDIs). 
\begin{figure}[!t]
	\centering
	\includegraphics[width=1\linewidth]{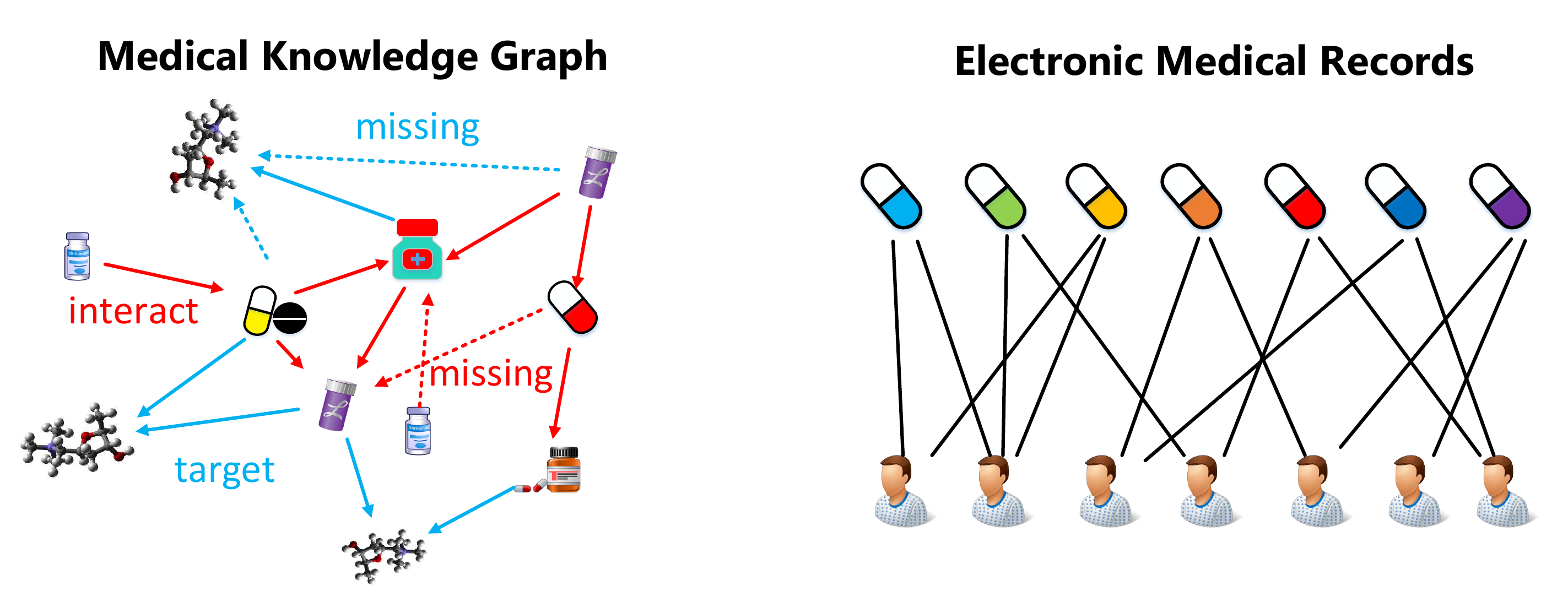}
	\vspace{-0.6cm}
	\caption{Left part is a medical knowledge graph with missing relationships. Right part is prescription records in EMRs, and each edge indicates that a patient takes a medicine. There is no relationship between medicines in EMRs.}
	\label{gap}
\end{figure}

As shown in Figure\ref{gap}, linking EMRs and medical knowledge graphs to generate a large and high-quality heterogeneous graph is a promising pathway for medicine recommendations in a wider scope, but never easy. Specifically, the newly designed system confronts with the following challenges: \textbf{1. Computational Efficiency}. Querying specialized medical entities and relationships based on conventional graph-based algorithms have limitations in portability and scalability. The computational complexity becomes unfeasible when the heterogeneous graph reaches a very large scale. \textbf{2. Data Incompleteness}. The medical knowledge graphs also follow the long-tail distribution as same as other types of large-scale knowledge bases. Data incompleteness is another serious problem existing among entities and relationships in such a distribution. For example, since the DDIs is not usually identified in the clinical trial phase, there is a lack of significant DDIs in DrugBank which cannot support a comprehensive precaution to the medication. Last but not least, medicine recommendation suffers from \textbf{3. Cold Start}. As conventional systems normally recommend medicines based on the historical records, the pace of recommendation changes cannot keep up with the frequent updates of new therapies and treatments in medical practices. Little information on adverse reactions to the newly updated medicines in historical EMRs or even in well-built knowledge graphs makes the evidence-based recommendation model hardly support the new medicines as updated recommendations.

Taking all the challenges above into account, we propose a novel medicine recommendation framework based on graph embedding techniques, inspired by the idea of link prediction. We name our framework as Safe Medicine Recommendation (SMR) throughout this paper. The recommendation process mainly includes:
\begin{enumerate}
	\item A large heterogeneous graph is constructed from EMRs and medical knowledge graphs, where the nodes are entities (medicines, diseases, patients), and the edges (or links) represent various relations between entities, such as drug-drug interactions. 
	
	\item The different parts of the generated heterogeneous graph (patient-medicine bipartite graph, patient-disease bipartite graph, medicine knowledge graph, disease knowledge graph) are embedded into a shared low-dimension space based on graph-based embedding models. Afterward, a joint learning algorithm is proposed to optimize the integrated graph simultaneously.
	
	\item Based on the learned embeddings, a new patient, represented by the vectors of his/her diagnoses, is modeled as an entity in the disease-patient graph. Recommending medicines for the patient is translated to predict links from the patient to medicines.
\end{enumerate}

The primary contributions of this work are summarized as follows:
\begin{enumerate}
	\item We have developed graph-based embedding models to learn the effective representations of patients, diseases, and medicines in a shared low-dimension space. The representation of medicines enables the proposed framework can even effectively recommend newly emerged medicines for patients, which distinguishes most of the existing works. 
	\item To recommend safe medicines for new patients, we propose a novel method for modeling a patient based on the learned graph embeddings and make a safe recommendation by minimizing the potential adverse drug reactions.
	\item We have conducted extensive experiments on large real-world datasets (MIMIC-III, DrugBank, and ICD-9 ontology) to evaluate the effectiveness of our framework. The experimental results have shown that the proposed framework outperforms all the compared methods.
	\item To our best knowledge, we firstly propose a framework to conduct Safe Medicine Recommendation (SMR) and formulate it as a link prediction problem. The implementation generates a high-quality heterogeneous graph in which relationships among patients, diseases, and medicines can be unveiled in a wider scope.  
\end{enumerate}

The remainder of this paper is organized as follows: Section 2 details our proposed framework SMR. Section 3 reports the experimental results and Section 4 reviews related work. Section 5 presents the conclusions and future work.

\section{The Proposed Framework}
In this section, we will first describe the notations and formulate medicine recommendation problem, and then present graph embedding models and how to use learned embeddings to recommend safe medicines for patients.

\begin{figure*}[!t]
	\centering
	\includegraphics[width=0.85\linewidth]{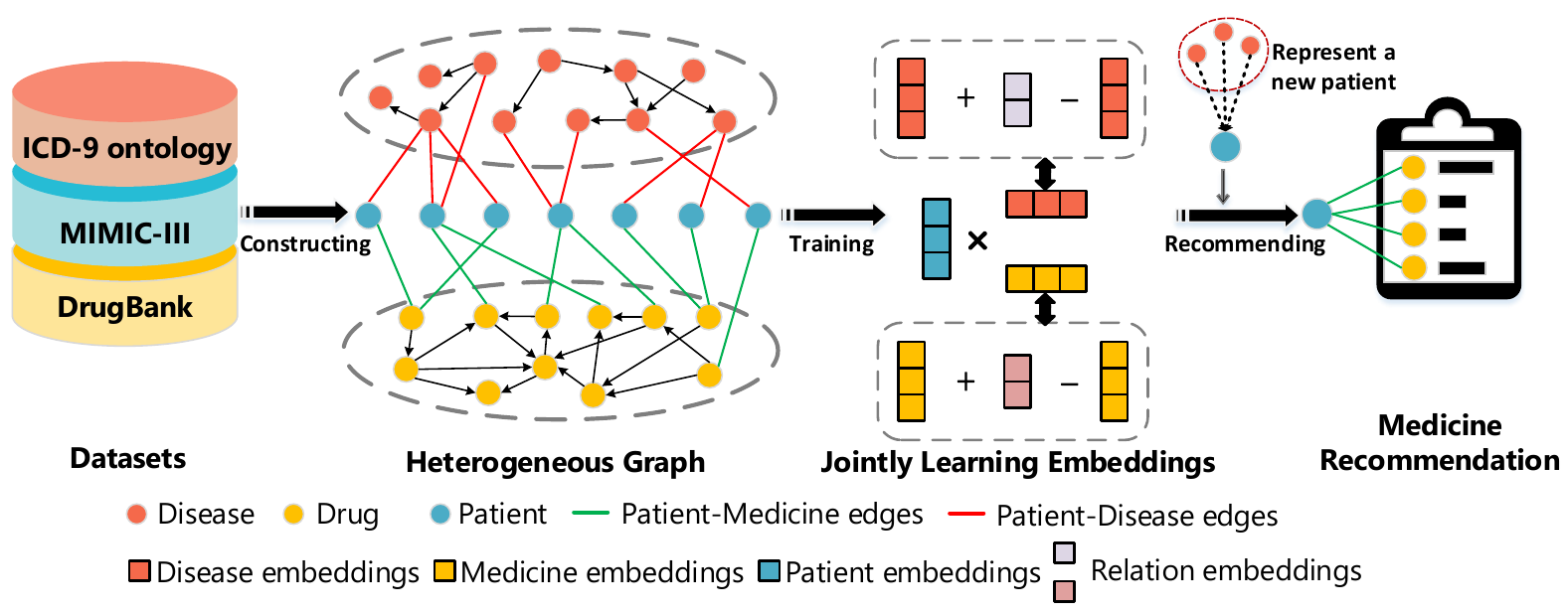}
	\vspace{-0.3cm}
	\caption{Overview of our framework. Patient-disease and patient-medicine graphs are all bipartite graphs, while disease and medicine graphs are general graphs. Patient, disease, and medicine are encoded into a low dimensional metric by graph-based methods. Diverse factors can be connected through patients.}
	\label{overview}
\end{figure*} 

\subsection{Problem Formulation}
Before we focus on the medicine recommendation problem, we first briefly introduce the important notations employed in the remainder of this paper. Table \ref{notation} also summarizes them.

\begin{table}
	\begin{tabular}{c|c}
		\hline
		Variable & Interpretation \\
		\hline
		$\mathcal{N}$, $\mathcal{R}$ & the set of entities and relations \\
		\hline
		$(h,r,t)$ & a triple in knowledge graph \\
		\hline
		$\mathbf{h}, \mathbf{t}$ & both are $k$ dimensional vector embeddings \\
		
		\hline
		$\mathbf{r}$ & a $d$ dimensional relation embedding \\
		\hline
		$\mathbf{H}_r$ & \tabincell{c}{a $k\times d$ dimensional projection matrix}\\
		
		\hline
		$p,m,d$ & a patient, medicine, disease \\
		\hline
		$\mathbf{p}$, $\mathbf{m}$, $\mathbf{d}$ & \tabincell{c}{ a $k$ dimensional vector of a patient, \\ a medicine or a disease } \\
		
		\hline
		$\mathbb{R}^k, \mathbb{R}^d$ &  \tabincell{c}{$k$ and $d$ dimensional latent space of \\ entities and relations}\\
		\hline
	\end{tabular}
	\caption{Notations.}
	\vspace{-0.4cm}
	\label{notation}
\end{table}

The medical knowledge graph describes the medical entities collected from the integrated sources, as well as relationships among these entities. For instance, a triple (\textit{glucocorticoid}, \textit{adverse interaction}, \textit{aspirin}) indicates that there is a relationship \textit{adverse interaction} from \textit{glucocorticoid} to \textit{aspirin} in DrugBank. We define the medical knowledge graph as follow.

\begin{myDef}[Medical Knowledge Graph]
	The medical knowledge graph $G=(\mathcal{N,R})$ is a set of triples in the form $(h,r,t)$, where $\mathcal{N}$ is a set of entities, $\mathcal{R}$ is a set of relations, $h,t\in \mathcal{N}$ and $r\in \mathcal{R}$.
\end{myDef}

To capture the co-relationships of patients, diseases, and medicines in EMRs, we define the patient-disease, patient-medicine bipartite graphs as follow.

\begin{myDef}[Patient-Medicine Bipartite Graph]
	The patient-medicine bipartite graph is denoted as $G_{pm}=(\mathcal{P}\cup \mathcal{M}, \mathcal{E}_{pm})$, where $\mathcal{P}$ is a set of patients and $\mathcal{M}$ is a set of medicines.  $\mathcal{E}_{pm}$ is the set of edges. If a patient $p_i$ takes a medicine $m_j$, there will be an edge $e_{ij}$ between them, otherwise none. The weight $w_{ij}$ of the edge between patient $p_i$ and medicine $m_j$ is defined as the	total times of patient $p_i$ takes the medicine $m_j$.
\end{myDef}

\begin{myDef}[Patient-Disease Bipartite Graph]
	The patient-disease bipartite graph is denoted as $G_{pd}=(\mathcal{P}\cup \mathcal{D}, \mathcal{E}_{pd})$, where $\mathcal{P}$ is a set of patients and $\mathcal{D}$ is a set of diseases. $\mathcal{E}_{pd}$ is the set of edges. If a patient $p_i$ is diagnosed with a disease $d_j$, there will be an edge $e_{ij}$ between them, otherwise none. The weight $w_{ij}$ is set to 1 when the edge $e_{ij}$ exists.
\end{myDef}
Figure \ref{overview} illustrates a heterogeneous graph by constructing patient-disease, patient-medicine bipartite graphs from MIMIC-III, and linking them to medical knowledge graphs, ICD-9 ontology, and DrugBank. Finally, we formally define the safe medicine recommendation problem as follows.
\begin{myPro}[Safe Medicine Recommendation] Given a patient $p$ and his/her diagnoses dataset $\mathcal{D}_p$, recommending safe medicines for each $d \in \mathcal{D}_p$ is predicting edges from $p$ to medicines dataset $\mathcal{M}$. The output is a set of medicines $\mathcal{M}_p$ with minimum drug-drug interactions.
\end{myPro}

\subsection{Model Description and Optimization}
In this section, we propose embedding learning approaches to encode the heterogeneous graph in the latent space and its optimization method.
\subsubsection{Medical Knowledge Graph Embedding}

A medical knowledge graph $G=(\mathcal{N,R})$ is a multi-relational graph, in which entities $\mathcal{N}$ and relations $\mathcal{R}$ can be different types. For a triple, $(h,r,t) \in G$, we use bold letters \textbf{h}, \textbf{r}, \textbf{t} to denote the corresponding embedding representations of $h, r, t$. Plenty of graph embedding methods has been proposed to encode a multi-relational graphs into a continuous vector space. Translation-based models \cite{bordes2013translating,wang2014knowledge,lin2015learning} regard the relation $r$ in each $(h, r, t)$ as a translation from $h$ to $t$ within the low dimensional space, i.e., $\mathbf{h+r-t}$, and perform much more effectively and efficiently than conventional models. TransR \cite{lin2015learning} is a state-of-the-art translation-based embedding approach. It represents entities and relations in distinct vector space bridged by relation-specific matrices to get better graph representations.

Consider the above reason, we set entities embeddings $\mathbf{h}, \mathbf{t} \in \mathbb{R}^k$ and relations embeddings $\mathbf{r} \in \mathbb{R}^d$. And we set a projection matrix $\mathbf{H}_r \in \mathbb{R}^{k\times d}$, which projects entities from entity space to relation space. We define the translations between entities and get the energy function $z(h,r,t)$ as:
\begin{equation}
	z(\mathbf{h,r,t})=b-\lVert \mathbf{hH}_r+\mathbf{r}-\mathbf{tH}_r \rVert_{L1/L2}
\end{equation}
where $b$ is a bias constant. 

Then, the conditional probability of a triple $(h,r,t)$ is defined as follows:
\begin{equation}
	P(h|r,t)=\frac{exp\{z(\mathbf{h,r,t})\}}{\sum_{\hat{h}\in \mathcal{N}}exp\{z(\mathbf{\hat{h},r,t})\}}
\end{equation}
and $P(t|h,r)$, $P(r|h,t)$ can be defined in the analogous manner. We define the likelihood of observing a triple  $(h,r,t)$ as:
\begin{equation}
	\begin{split}
		\mathcal{L}(h,r,t)=logP(h|r,t)&+logP(t|h,r)\\&+logP(r|h,t)
	\end{split}
	\label{hrt}
\end{equation}
We define an objective function by maximizing the conditional likelihoods of existing triples in $G$:
\begin{equation}
	\mathcal{L}_G=\sum\limits_{(h,r,t)\in G} \mathcal{L}(h,r,t)
	\label{lg}
\end{equation}

Based on Eq.(\ref{lg}), the objective functions of medicine and disease knowledge graph $G_m=(\mathcal{N}_m,\mathcal{R}_m)$, $G_d=(\mathcal{N}_d,\mathcal{R}_d)$ can be defined respectively:
\begin{equation}
	\mathcal{L}_{G_m}=\sum\limits_{(h_m,r_m,t_m)\in G_m} \mathcal{L}(h_m,r_m,t_m)
	\label{lm}
\end{equation}

\begin{equation}
	\mathcal{L}_{G_d}=\sum\limits_{(h_d,r_d,t_d)\in G_d} \mathcal{L}(h_d,r_d,t_d)
	\label{ld}
\end{equation}

\subsubsection{Bipartite Graph Embedding}
Different from the medical knowledge graph, the patient-disease, patient-medicine are bipartite graphs. A bipartite graph has only one single type of relations. For a bipartite graph, LINE \cite{tang2015line} model achieves the state-of-the-art performance of encoding the entities into a continuous vector space while preserving co-relations information of the graph. Hence, we follow LINE and set patients, medicines, and diseases embeddings $\mathbf{p}, \mathbf{m}, \mathbf{d} \in \mathbb{R}^k$. We present the process of encoding patient-medicine bipartite graph as follow.


Given a patient-medicine bipartite graph $G_{pm}=(\mathcal{P}\cup \mathcal{M}, \mathcal{E}_{pm})$. We first define the conditional probability of that a patient $p_i$ in set $\mathcal{P}$ takes medicine $m_j$ in set $\mathcal{M}$ as follow:

\begin{equation}
	P(m_j|p_i) =  \frac{exp\{z(\mathbf{p}_i,\mathbf{m}_j)\}}{\sum_{\hat{m}_j\in \mathcal{P}_i}exp\{z(\mathbf{p}_i,\mathbf{\hat{m}}_j)\}}
	\label{condi}
\end{equation}
where $z(\mathbf{p}_i,\mathbf{m}_j) = \mathbf{m}_j^T\cdot \mathbf{p}_i$, $\mathbf{p}_i$ is the embedding vector of the patient $p_i$ in $\mathcal{P}$, and $\mathbf{m}_j$ is the embedding vector of medicine $m_j$ in $\mathcal{M}$. Eq. (\ref{condi}) defines a conditional distribution $P(\cdot|p_i)$ over all medicines in $\mathcal{M}$. The empirical distribution $\hat{P}(\cdot|p_i)$is defined as $\hat{P}(m_j|p_i) = \frac{w_{ij}}{sum_i}$, where $w_{ij}$ is the weight of the edge $e_{ij}$ and $sum_i=\sum_{j}{w_{ij}}$ is the total times that the patient $p_i$ takes medicines. We maximize the following objective function:
\begin{equation}
	\mathcal{L}_{G_{pm}} = -\sum \limits_{p_i \in \mathbf{P}} \lambda_i d(\hat{P}(\cdot|p_i),P(\cdot|p_i))
\end{equation}
where $d(\cdot, \cdot)$  is the distance between two distributions. In this paper, we use KL-divergence to compute $d(\cdot, \cdot)$. As $sum_i$ is different from patients, we use $\lambda_i=sum_i$ in the objective function to represent the personalization of the patient $p_i$ in the graph. After omitting some constants, we have:
\begin{equation}
	\mathcal{L}_{G_{pm}} = \sum \limits_{e_{ij} \in \mathcal{E}_{pm}} w_{ij} \log(P(m_j|p_i))
	\label{lpm}
\end{equation}
For the patient-disease bipartite graph, we can get the object function $\mathcal{L}_{G_{pd}}$ in the analogous manner:
\begin{equation}
	\mathcal{L}_{G_{pd}} = \sum \limits_{e_{ij} \in \mathcal{E}_{pd}} w_{ij} \log(P(d_j|p_i))
	\label{lpd}
\end{equation}

\subsubsection{Optimization and Training}
To learn the medical knowledge graph and bipartite graphs embeddings simultaneously, an intuitive approach is to collectively embed the four graphs (patient-medicine bipartite graph, patient-disease bipartite graph, medicine knowledge graph, disease knowledge graph) by maximizing the sum of the four logarithm likelihood objective functions just as follow:
\begin{equation}
	\mathcal{L}(X) = \mathcal{L}_{G_m} + \mathcal{L}_{G_d} + \mathcal{L}_{G_{pm}} + \mathcal{L}_{G_{pd}}+\gamma C(X)
	\label{of}
\end{equation}
where $X$ stands for the embeddings $\mathbb{R}_k$, $\mathbb{R}_d$ of entities and relations in the heterogeneous graph we construct, $\gamma$ is a hyper-parameter weighting the regularization factor $C(X)$, which is defined as follows:
\begin{equation}
	\begin{split}
		&C(X) =  \sum_{n_m \in \mathcal{N}_m}[||n_m||-1]_+ +\sum_{n_d \in \mathcal{N}_d}[||n_d||-1]_+\\
		&\qquad \quad+ \sum_{r_m \in \mathcal{R}_m}[||r_m||-1]_+ + \sum_{r_d \in \mathcal{R}_d}[||r_d||-1]_+\\
		& +\sum_{p \in \mathcal{P}}[||p||-1]_+ +\sum_{d \in \mathcal{D}}[||d||-1]_+ +\sum_{m \in \mathcal{M}}[||m||-1]_+
	\end{split}
\end{equation} 
where $[x]_+ = max(0,x)$ denotes the positive part of $x$. The regularization factor will normalize the embeddings during learning. And we adopt the asynchronous stochastic gradient algorithm (ASGD) \cite{recht2011hogwild} to maximize the transformed objective function.

Optimizing objective functions Eq. (\ref{lm}), Eq. (\ref{ld}), Eq. (\ref{lpm}) and Eq. (\ref{lpd}) in Eq.(\ref{of}) are computationally expensive, as calculating them need to sum over the entire set of entities and relations. To address this problem, we use the negative sampling method \cite{mikolov2013distributed} to transform the objective functions. 

For Eq.(\ref{lm}) and Eq.(\ref{ld}), we should transform $logP(t|h,r)$, $logP(r|h,t)$,$logP(h|r,t)$ in Eq.(\ref{hrt}). Taking $P(t|h,r)$ as an example, we maximize the following objective function instead of it:
\begin{equation}
	\begin{split}
		&\log \sigma(z(\mathbf{h},\mathbf{r},\mathbf{t}))\\+
		&\sum_{n=1}^{C_1} E_{\widetilde{h}_n\sim z_{neg}(\{(\widetilde{h},r,t)\})}[\sigma(z(\widetilde{\mathbf{h}}_n,\mathbf{r},\mathbf{t}))]
		\label{okg}
	\end{split}
\end{equation}
where $C_1$ is the number of negative examples, $\sigma(x)=1/(1+exp(-x))$ is the sigmoid function. $\{(\widetilde{h},r,t)\}$ is the invalid triple set, and $z_{neg}$ is a function randomly sampling instances from $\{(\widetilde{h},r,t)\}$. When a positive triple $(h, r, t) \in G$ is selected, to maximize Eq.(\ref{okg}), $C_1$ negative triples are constructed by sampling entities from an uniform distribution over $\mathcal{N}$ and replacing the head of $(h, r, t)$. The transformed objective of
$log P(r|h, t)$, $log P(t|h, r)$ are maximized in the same manner,
but for $log P(r|h, t)$, the negative relations are sampled from a uniform distribution over
$\mathcal{R}$ to corrupt the positive relation $r\in (h, r, t)$. We iteratively select random mini-batch from the training set to learn embeddings until converge.

For Eq. (\ref{lpm}), we also use the negative sampling method to transform it to the following objective function:
\begin{equation}
	\begin{split}
		&\log \sigma(z(\mathbf{p}_i,\mathbf{m}_j))\\&+\sum_{n=1}^{C_2} E_{\widetilde{m}_n\sim z_{neg}(\widetilde{m})}[\log \sigma(z({\mathbf{p}}_i,\widetilde{\mathbf{m}}_n))]
		\label{optm}
	\end{split}
\end{equation}
where $\sigma(x)=1/(1+exp(-x))$ is the sigmoid function, $C_2$ is the number of negative edges. $z_{neg}(\widetilde{m})\propto sum_{\widetilde{m}}^{3/4}$ according to the empirical setting of \cite{mikolov2013distributed}, $sum_{\widetilde{m}}$ is the total number of times that the medicine $\widetilde{m}$ is taken by patients. 
we can simplify Eq.(\ref{lpd}) and maximize it in the same way.

Finally, we can efficiently learn the embeddings of different types of parts in the heterogeneous graph. 

\subsection{Safe Medicine Recommendation Process}
In this section, we present how to recommend safe medicines based on the learned embeddings and diagnoses of a given patient. For an existing patient $p$, we use the learned embedding $\mathbf{p}$ to predict new medicine recommendations. For a new given patient $p$, we first use diseases embeddings of $p$'s diagnoses to represent $\mathbf{p}$, and then recommend safe medicines for $p$, as shown in Figure \ref{overview},.

\subsubsection{New Patient Model}
We aim to present a new patient $p$ by his/her diagnoses embeddings. We should consider the time sequence of diseases that a patient is diagnosed, especially for the patient with multiple diseases. Assume a patient $p$ in the hospital or on medication is associated with $n$ ranked diseases according to their timestamps in an increasing order. Then, the patient embedding $\mathbf{p}$ can be encoded as follow:
\begin{equation}
	\mathbf{p}=\sum_{t=1}^n exp^{-t} \cdot \mathbf{d}_t \label{patient}
\end{equation}
where $ \mathbf{d}_t$ is the $t$-th embedding of disease $ d_t$.

\subsubsection{Medicine Recommendation}
Given a query patient $p$ with the query disease $d$, i.e., $q=(p, d)$, we first project disease $d$ and patient $p$ into their latent space, and then select top-$k$ safe medicines\footnote{In MIMIC-III, patients in ICUs are sicker and usually need more medicines for a diagnosis. We set $k$=3 in this paper.}. More precisely, given a query $q=(p, d)$, for each medicine $m$ which could be useful for $p$, we compute its ranking score as in Eq. (\ref{sf}), and then select the medicine $m$ with the top-$k$ highest ranking scores as the recommendation.
\begin{equation}
	S(q, m_n)=\mathbf{p}^T\cdot  \mathbf{m}_n  - \sum_{o=1}^{n-1}\lVert \mathbf{m}_n+ \mathbf{r}_{interation}-\mathbf{m}_o \rVert_{L1/L2} \label{sf}
\end{equation}

where $ \mathbf{p}$ is the representation of patient $p$ and $m_n$ is the $n$-th medicines to be considered from medicines $\mathcal{M}$ based on the the already selected medicine $m_1,...,m_{n-1}$. 

\section{Experiments and Evaluation}
We attempt to demonstrate the effectiveness of our recommendation method in this section, which is referred to as SMR in this paper. In particular, we expect to answer ``\textbf{how well does our method compare with the competing techniques?}" in Section 3.2. The results show that our recommendation method significantly outperforms the three baselines. The detailed experimental settings of our evaluations are described in Section 3.1.

\subsection{Experimental Settings}

\subsubsection{Data Sets}Our experiments are performed on the real EMRs datasets, MIMIC-III \cite{johnson2016mimic}, and two medical knowledge graphs, ICD-9 ontology \cite{schriml2011disease} and DrugBank \cite{drugbank}. These real datasets are publicly available in different forms.
\begin{table}  
	\small{  
		\begin{tabular}{lr|lr}
			\hline
			\multicolumn{2}{c|}{\textbf{Entities} } & \multicolumn{2}{c}{\textbf{Relations} }\\
			\hline
			\#Disease &6,985 &\#Medicine-related & 71,460\\
			
			\#Medicine &8,054 & \#Disease-related& 170,579\\
			\#Patient & 46,520 & \#Patient-Disease& 489,923\\
			\#Medicine-related & 305,642&  \#Patient-Medicine& 975,647\\
			
			\hline
		\end{tabular}
		\caption{Entities and relations in the heterogeneous graph.} 
		\label{tab:er}  
	}
\end{table}	

\begin{itemize}
	\item MIMIC-III (Medical Information Mart for Intensive Care III) collected bedside monitor trends, electronic medical notes, laboratory test results, and waveforms from the ICUs (Intensive Care Units) of Beth Israel Deaconess Medical Center between 2001 and 2012. It contains distinct 46,520 patients, 650,987 diagnoses and 1,517,702 prescription records that associated with 6,985 distinct diseases and 4,525 medicines. 
	
	\item ICD-9 ontology\footnote{\url{http://bioportal.bioontology.org/ontologies/ICD9CM}} (International classification of diseases-version 9) contains 13,000 international standard codes of diagnoses and the relationships between them.
	
	\item DrugBank is a bioinformatics/cheminformatics resource which consists of medicine related entities. The medical knowledge graph version\footnote{\url{http://wifo5-03.informatik.uni-mannheim.de/drugbank/}} contains 8,054 medicines, 4,038 other related entities (e.g., protein or drug targets) and 21 relationships. 
\end{itemize}

\begin{table}  
	\begin{floatrow}  
		\capbtabbox{  
			\begin{tabular}{c|c|r}
				\hline
				& Prediction accuracy & DDIs rate \\
				\hline
				Rule-based & 0.3068 & 32.01\%  \\
				\hline
				$K$-Most frequent& 0.3473 &14.08\%\\
				\hline
				LEAP &  0.5582 & 1.66\% \\
				\hline
				SMR & \textbf{0.6113} & \textbf{0.17\%} \\
				\hline
			\end{tabular}  
		}{  
			\caption{Experiments on medicine group 1.}
			\label{tab:ra}    
		}  
	\end{floatrow}  
\end{table}

\begin{table}  
	\begin{floatrow} 
		\capbtabbox{ 
			\begin{tabular}{c|c|c}
				\hline
				& Prediction accuracy & DDIs rate \\
				\hline
				Rule-based &0.2736& 27.01\%\\
				\hline
				K-Most frequent & NA &NA\\
				\hline
				LEAP & NA &NA\\
				\hline
				SMR & \textbf{0.5214}&\textbf{2.01\%}\\
				\hline
			\end{tabular}  
		}{  
			\caption{Experiments on medicine group 2.}  
			\label{tab:adi}
		}  
	\end{floatrow}  
\end{table}  
\noindent\textbf{Heterogeneous Graph Construction}
We connect MIMIC-III, ICD-9 ontology, and DrugBank (medicine group 1) by constructing the patient-medicine bipartite graph and the patient-disease bipartite graph.

For the patient-disease bipartite graph, MIMIC-III provides ICD-9 codes for diagnoses, which implicitly the diagnoses of MIMIC-III can be linked to ICD-9 ontology by string matching. For the patient-medicine bipartite graph, the prescriptions in MIMIC-III consist of the drug information, e.g., the names, the duration, and the dosage. However, various names to a single type of medicine in MIMIC-III exist due to some noisy words (20\%, 50ml, glass bottle, etc.), which becomes an obstacle to link medicine names to DrugBank when directly applying to the string matching method. We use an entity linking method \cite{wang2017pdd} instead to address this problem. Table \ref{tab:er} shows the statistic of the heterogeneous graph we construct. The heterogeneous graph will be used to learn low-dimension representations of entities and relations by the SMR framework. Afterward, we categorize the medicines in the heterogeneous graph into two groups: 1). The first group consists of all 4,525 medicines that are recorded in EMRs, and will be used as inputs of the baseline methods. 2). The second group contains 3,529 medicines that haven't been observed in EMRs, and will be used as test data for cold start recommendation.

\subsubsection{Baselines}
We compare our SMR with the following state of the art methods:
\begin{itemize}
	\item Rule-based method \cite{almirall2012designing} recommends medicines based on mappings from existing medicine categories to diseases in the MEDI database \cite{wei2013development}. For each disease, a drug is assigned to the patient according to the mappings.
	\item $K$-Most frequent method is a basic baseline which retrieves the top $K$ medicines that most frequently co-occur with each disease as their recommendation. We set $K=3$ in this paper.
	\item LEAP method \cite{zhang2017leap} uses a Multi-Instance Multi-Label learning framework to train a predictive model taking disease conditions as input features and yielding multiple medicine labels as recommendations.
\end{itemize}

\subsection{Evaluation Methods}
To guarantee medicine recommendations generated by SMR work effectively, we evaluate four indice, the prediction accuracy, the ability to avoid adverse drug-drug interactions, the experienced clinical doctor assessments, and the capacity to process cold start problem. In all experiments, the ratio of training to validation to test sets is 0.7:0.1:0.2. The hyper-parameters was adjusted by a validation set.

\subsubsection{Prediction Accuracy and DDIs Rate}
We utilize Jaccard Coefficient to compare the similarity of the prescriptions generated by SMR and the corresponding prescriptions written by doctors. Given the recommendation medicines set $M_i$ generated by SMR for a patient $p_i$, $\check{M}_i$ is the medicines set prescribed by doctors in the data. The mean of Jaccard coefficient is defined as follows, 
\begin{equation}
	Jaccard=\frac{1}{K} \sum_{i=1}^K \frac{|M_{i}\cap \check{M}_{i}|}{|M_{i}\cup \check{M}_{i}|}
\end{equation}
where $K$ is the number of samples in a test set. Table \ref{tab:ra} shows the accuracy of the baselines and SMR on medicine group 1, the rule-based method performs the worst because it is the only one provides a general recommendation for a specific diagnosis and it is not able to endow personalized recommendations, especially for the patients with multiple diseases. The frequency of each medicine-disease pair remains high in ICUs. Hence, recommendations based on frequency, $K$-Most frequent method, also work deficiently. Our method SMR outperforms LEAP by 1.49\% because more accurate medical facts are involved in medical knowledge graphs rather than the prescription information in EMRs.

We extract all adverse drug-drug interactions (DDIs) from DrugBank to evaluate whether medicine recommendations embrace unsafe DDIs. Table \ref{tab:ra} shows the percentages of different medicine recommendations consisting of adverse DDIs. The result indicates that SMR can recommend most harmless medicines for patients as its drug interaction rate is the lowest. The rule-based method and $K$-Most frequent method select medicines by a greedy strategy only regardless of specific adverse DDIs. For the rarely used medicines and unknown DDIs in EMRs,
SMR is more reliable than LEAP. The reason is that SMR can predict each patient-medicine link and compute potential hidden DDIs by the learned embeddings of medical knowledge graphs.

\subsubsection{Cold Start}
We evaluate the ability of baselines and SMR in addressing cold start medicine recommendations on the medicine group 2. The results are reported in Table \ref{tab:adi}. $K$-Most frequent method and LEAP are not applicable (NA) on recommending new medicines in the cold start scenario. Since our SMR process can present new medicines by the learned vector representations of used medicines, the potential patient-medicine links between cold start medicines and patients will be captured correspondingly. In other words, SMR can leverage not only the patient-medicine links in EMRs but also the medical knowledge graphs when recommending cold start medicines.

\begin{table}[t]
	\small{
		\centering
		\begin{tabular}{c|l|l}
			\hline
			Diagnosis & Methods & Medicine Recommendations\\
			\hline
			\multirow{4}{3.5cm}{Sepsis\\ Acute respiratry failure \\Hypertension} & Rule-based &  \tabincell{c} {Teicoplanin, Metoprolol} \\ 
			\cline{2-3}
			& $K$-Most frequent &\tabincell{l}{ Vancomycin, Furosemide,\\ Metoprolol, Insulin}\\
			\cline{2-3}
			& LEAP &\tabincell{l}{Vancomycin, Furosemide,\\ Metoprolol Tartrate}\\
			\cline{2-3}
			& SMR & \tabincell{l}{Vancomycin, Furosemide, \\Amlodipine, Norepinephrine, Acetaminophen}\\
			\hline
			\multirow{4}{3.5cm}{Type 2 diabetes\\ Rheumatoid arthritis \\Hypertension \\ Hyperlipidemia} & Rule-based & \tabincell{l}{Gliclazide, Phenylbutazone, \\Sulfasalazine, Fenofibrate}   \\
			\cline{2-3}
			& $K$-Most frequent & \tabincell{l}{Furosemide, Tolbutamide,\\ Phenylbutazone, Metoprolol, \\Insulin, Acetaminophen}\\
			\cline{2-3}
			& LEAP & \tabincell{l}{Metformin, Amethopterin, \\Amiloride/HCTZ, Fenofibrate}\\
			\cline{2-3}
			& SMR & \tabincell{l}{ Metformin, Insulin, \\Acetaminophen, Nifedipine, Fenofibrate}\\
			\hline
		\end{tabular}
	} 
	\caption{Examples of medicine recommendations generated by Baselines and SMR.}  
	\label{tab:cs}
\end{table}

\subsubsection{Clinical Assessment}
\begin{figure}[!t]
	\centering
	\includegraphics[width=0.6\linewidth]{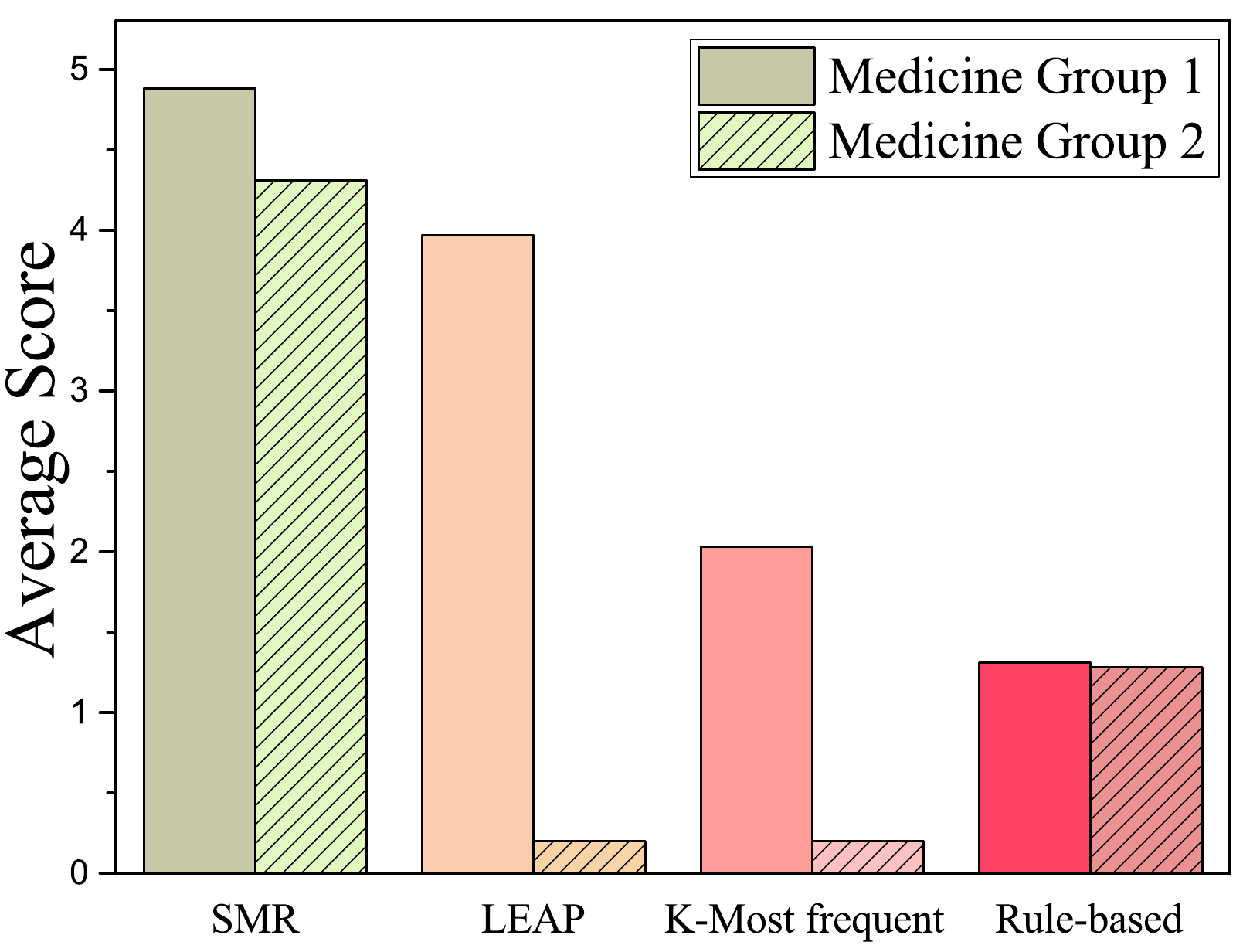}
	\caption{Clinical Assessment.}
	\label{score}
\end{figure}

We invited three experienced clinical experts to evaluate the effectiveness of the medicine recommendations by scoring on a 6-point scale: 5 corresponding to completely cover all diagnoses without DDIs; 4 to partially (at least 50\%) diagnoses include without DDIs; 3 to completely cover all diagnoses with DDIs; 2 to less than 50\% diagnoses without DDIs; 1 to partially (at least 50\%) diagnoses covered with DDIs; 0 to less than 50\% diagnoses with DDIs. The average score of three experts is used as the final clinical assessment score for each recommendation, as shown in Figure \ref{score}.

\subsubsection{Case Study}
In table \ref{tab:cs} we illustrate two events of medicine recommendations on medicine group 1 for patients associated with multiple types of diseases. SMR is qualified to succeed in all these two cases when comparing it against other baselines. For the first patient, SMR recommended a set of medicines with 100\% coverage, with Vancomycin for Sepsis, Norepinephrine, Acetaminophen for respiratry failure, Furosemide and Amlodipine for Hypertension. In contrast, other baselines are not capable of make an adequate consideration. The rule-based method adopted Teicoplanin, targeting Sepsis only and not appropriate. The $K$-Most frequent method and LEAP only selected Vancomycin for Sepsis and other medicines for Hypertension. For the second patient, SMR recommends more suitable medicines than LEAP and Rule-based method, i.e., Metformin and Insulin for Type 2 diabetes, Acetaminophen to release Rheumatoid arthritis, Nifedipine for Hypertension, and Fenofibrate for Hyperlipidemia. There is an adverse DDI among the medicines recommended by the $K$-Most frequent method. Tolbutamide and Phenylbutazone can lead to harmful, potentially fatal effects when taken together. This case also indicates SMR can avoid the adverse DDIs when recommending medicines.

\section{Related Work}
In this section, we discuss related work, including medicine recommendation, medical knowledge graphs and embeddings.

\subsubsection{Medicine Recommendation}
As introduced in Section 1, two types of methods, rule-based protocols \cite{chen2016physician,gunlicks2016pilot,almirall2012designing}, and supervised-learning-based methods \cite{zhang2017leap,zhang2014towards}, are currently utilizing EMRs to recommending medicines. Ideally, medicine recommendation systems aim to tailor treatment to the individual characteristics of each patient \cite{fernald2011bioinformatics}. Hence, medicine recommendation has also received attention recently in genetics/genomics research fields. There are already existing medicine recommendation systems \cite{rosen2008selecting,bennett2013artificial} by leveraging genetics/genomics information of patients in current practice, such information is not yet widely available in everyday clinical practice, and is insufficient since it only addresses one of many factors affecting response to medication. A previous pre-print version of our work \cite{wang2017safe} mainly introduced our idea and recommendation model, while in this paper we provide a more in-depth analysis of the experiment and a summary of the relevant work.

\subsubsection{Medical Knowledge Graphs and Embeddings}
Recent evaluation efforts on knowledge graphs have focused on automatic knowledge base population and completion. Some knowledge graphs have also been constructed from huge volumes of medical databases over the last years, such as \cite{ernst2015knowlife}, Bio2RDF\cite{dumontier2014bio2rdf}, and Chem2Bio2RDF\cite{chen2010chem2bio2rdf}. Medical knowledge graphs contain an abundance of basic medical facts of medicines and diseases and provide a pathway for medical discovery and applications, such as effective medicine recommendations. Unfortunately, such medical knowledge graphs suffer from serious data incomplete problem, which impedes its application in the field of clinical medicine. Recently, Celebi et al.  \cite{celebi2019evaluation} proposed a knowledge graph embedding approach for drug-drug interaction prediction in realistic scenario. MedGraph \cite{hettige2019medgraph} is a new EMR embedding framework which introduces a graph-based data structure to capture both structural visit-code co-location information structurally and temporal visit sequencing information. However, they only focused on the real-world biomedical embeddings learning and do not directly harness the medicine recommendation task, thus the specific clinical medicine information is not sufficiently tailored. In contrast, our model jointly embeds diseases, medicines, patients, and uses the learned embeddings to decompose the medicine recommendation into a linked prediction process while considering the patient’s diagnoses and adverse drug reactions.

\section{Conclusion and Future Work}
In this paper, we propose a novel framework SMR to recommend safe medicines for patients, especially for the patients with multiple diseases. SMR first constructs a high-quality heterogeneous graph by bridging EMRs (MIMIC-III) and medical knowledge graphs (ICD-9 ontology and DrugBank). Then, SMR jointly embeds diseases, medicines, patients, and their corresponding relations into a shared lower dimensional space. Finally, SMR uses the embeddings to decompose the medicine recommendation into a linked prediction process considering the clinical diagnoses and adverse DDI reactions. Extensive experiments on real world datasets are conducted and demonstrate the effectiveness of SMR. In future work, we will improve the linking accuracy by considering more information of patients, such as the clinical outcomes and demographics.

\section{Acknowledgment}
This work was supported by National Science Foundation of China with Grant Nos. 61906037 and U1736204; the Fundamental Research Funds for the Central Universities with Grant No.4009009106 and 22120200184; the CCF-Baidu Open Fund.

\bibliography{mybibfile}

\begin{thebibliography}{10}
\expandafter\ifx\csname url\endcsname\relax
  \def\url#1{\texttt{#1}}\fi
\expandafter\ifx\csname urlprefix\endcsname\relax\def\urlprefix{URL }\fi
\expandafter\ifx\csname href\endcsname\relax
  \def\href#1#2{#2} \def\path#1{#1}\fi

\bibitem{chen2016physician}
Z.~Chen, K.~Marple, E.~Salazar, G.~Gupta, L.~Tamil, A physician advisory system
  for chronic heart failure management based on knowledge patterns, Theory and
  Practice of Logic Programming 16~(5-6) (2016) 604--618.

\bibitem{almirall2012designing}
D.~Almirall, S.~N. Compton, M.~Gunlicks-Stoessel, N.~Duan, S.~A. Murphy,
  Designing a pilot sequential multiple assignment randomized trial for
  developing an adaptive treatment strategy, Statistics in Medicine 31~(17)
  (2012) 1887--1902.

\bibitem{zhang2017leap}
Y.~Zhang, R.~Chen, J.~Tang, W.~F. Stewart, J.~Sun, Leap: Learning to prescribe
  effective and safe treatment combinations for multimorbidity, in: Proceedings
  of the 23rd ACM SIGKDD International Conference on Knowledge Discovery and
  Data Mining, ACM, 2017, pp. 1315--1324.

\bibitem{panagioti2015multimorbidity}
M.~Panagioti, J.~Stokes, A.~Esmail, P.~Coventry, S.~Cheraghi-Sohi, R.~Alam,
  P.~Bower, Multimorbidity and patient safety incidents in primary care: a
  systematic review and meta-analysis, PloS One 10~(8) (2015) e0135947.

\bibitem{elderlypatients}
D.~N. Juurlink, M.~Mamdani, A.~Kopp, A.~Laupacis, D.~A. Redelmeier, Drug-drug
  interactions among elderly patients hospitalized for drug toxicity, Journal
  of The American Medical Association 289~(13) (2003) 1652--1658.

\bibitem{adverse}
I.~R. Edwards, J.~K. Aronson, Adverse drug reactions: definitions, diagnosis,
  and management, The Lancet 356~(9237) (2000) 1255--1259.

\bibitem{35errors}
L.~L. Leape, D.~W. Bates, D.~J. Cullen, J.~Cooper, H.~J. Demonaco, T.~Gallivan,
  R.~Hallisey, J.~Ives, N.~Laird, G.~Laffel, et~al., Systems analysis of
  adverse drug events, Journal of The American Medical Association 274~(1)
  (1995) 35--43.

\bibitem{drugbank}
V.~Law, C.~Knox, Y.~Djoumbou, T.~Jewison, A.~C. Guo, Y.~Liu, A.~Maciejewski,
  D.~Arndt, M.~Wilson, V.~Neveu, et~al., Drugbank 4.0: shedding new light on
  drug metabolism, Nucleic Acids Research 42~(D1) (2014) D1091--D1097.

\bibitem{schriml2011disease}
L.~M. Schriml, C.~Arze, S.~Nadendla, Y.-W.~W. Chang, M.~Mazaitis, V.~Felix,
  G.~Feng, W.~A. Kibbe, Disease ontology: a backbone for disease semantic
  integration, Nucleic Acids Research 40~(D1) (2011) D940--D946.

\bibitem{bordes2013translating}
A.~Bordes, N.~Usunier, A.~Garcia-Duran, J.~Weston, O.~Yakhnenko, Translating
  embeddings for modeling multi-relational data, in: Advances in neural
  information processing systems, 2013, pp. 2787--2795.

\bibitem{wang2014knowledge}
Z.~Wang, J.~Zhang, J.~Feng, Z.~Chen, Knowledge graph embedding by translating
  on hyperplanes., in: Proceedings of the 28th AAAI Conference on Artificial
  Intelligence, 2014, pp. 1112--1119.

\bibitem{lin2015learning}
Y.~Lin, Z.~Liu, M.~Sun, Y.~Liu, X.~Zhu, Learning entity and relation embeddings
  for knowledge graph completion., in: Proceedings of the 29th AAAI Conference
  on Artificial Intelligence, 2015, pp. 2181--2187.

\bibitem{tang2015line}
J.~Tang, M.~Qu, M.~Wang, M.~Zhang, J.~Yan, Q.~Mei, Line: Large-scale
  information network embedding, in: Proceedings of the 24th International
  Conference on World Wide Web, International World Wide Web Conferences
  Steering Committee, 2015, pp. 1067--1077.

\bibitem{recht2011hogwild}
B.~Recht, C.~Re, S.~Wright, F.~Niu, Hogwild: A lock-free approach to
  parallelizing stochastic gradient descent, in: Advances in Neural Information
  Processing Systems, 2011, pp. 693--701.

\bibitem{mikolov2013distributed}
T.~Mikolov, I.~Sutskever, K.~Chen, G.~S. Corrado, J.~Dean, Distributed
  representations of words and phrases and their compositionality, in: Advances
  in Neural Information Processing Systems, 2013, pp. 3111--3119.

\bibitem{johnson2016mimic}
A.~E. Johnson, T.~J. Pollard, L.~Shen, L.-w.~H. Lehman, M.~Feng, M.~Ghassemi,
  B.~Moody, P.~Szolovits, L.~A. Celi, R.~G. Mark, Mimic-iii, a freely
  accessible critical care database, Scientific Data 3.

\bibitem{wang2017pdd}
M.~Wang, J.~Zhang, J.~Liu, W.~Hu, S.~Wang, X.~Li, W.~Liu, Pdd graph: Bridging
  electronic medical records and biomedical knowledge graphs via entity
  linking, in: International Semantic Web Conference, Springer, 2017.

\bibitem{wei2013development}
W.-Q. Wei, R.~M. Cronin, H.~Xu, T.~A. Lasko, L.~Bastarache, J.~C. Denny,
  Development and evaluation of an ensemble resource linking medications to
  their indications, Journal of the American Medical Informatics Association
  20~(5) (2013) 954--961.

\bibitem{gunlicks2016pilot}
M.~Gunlicks-Stoessel, L.~Mufson, A.~Westervelt, D.~Almirall, A pilot smart for
  developing an adaptive treatment strategy for adolescent depression, Journal
  of Clinical Child \& Adolescent Psychology 45~(4) (2016) 480--494.

\bibitem{zhang2014towards}
P.~Zhang, F.~Wang, J.~Hu, R.~Sorrentino, Towards personalized medicine:
  leveraging patient similarity and drug similarity analytics, Proceedings of
  AMIA Summits on Translational Science 2014 (2014) 132.

\bibitem{fernald2011bioinformatics}
G.~H. Fernald, E.~Capriotti, R.~Daneshjou, K.~J. Karczewski, R.~B. Altman,
  Bioinformatics challenges for personalized medicine, Bioinformatics 27~(13)
  (2011) 1741--1748.

\bibitem{rosen2008selecting}
M.~Rosen-Zvi, A.~Altmann, M.~Prosperi, E.~Aharoni, H.~Neuvirth,
  A.~S{\"o}nnerborg, E.~Sch{\"u}lter, D.~Struck, Y.~Peres, F.~Incardona,
  et~al., Selecting anti-hiv therapies based on a variety of genomic and
  clinical factors, Bioinformatics 24~(13) (2008) i399--i406.

\bibitem{bennett2013artificial}
C.~C. Bennett, K.~Hauser, Artificial intelligence framework for simulating
  clinical decision-making: A markov decision process approach, Artificial
  Intelligence in Medicine 57~(1) (2013) 9--19.

\bibitem{wang2017safe}
M.~Wang, M.~Liu, J.~Liu, S.~Wang, G.~Long, B.~Qian, Safe medicine
  recommendation via medical knowledge graph embedding, arXiv preprint
  arXiv:1710.05980.

\bibitem{ernst2015knowlife}
P.~Ernst, A.~Siu, G.~Weikum, Knowlife: a versatile approach for constructing a
  large knowledge graph for biomedical sciences, BMC Bioinformatics 16~(1)
  (2015) 157.

\bibitem{dumontier2014bio2rdf}
M.~Dumontier, A.~Callahan, J.~Cruz-Toledo, P.~Ansell, V.~Emonet, F.~Belleau,
  A.~Droit, Bio2rdf release 3: a larger connected network of linked data for
  the life sciences, in: Proceedings of the 2014 International Conference on
  Posters \& Demonstrations Track-Volume 1272, CEUR-WS. org, 2014, pp.
  401--404.

\bibitem{chen2010chem2bio2rdf}
B.~Chen, X.~Dong, D.~Jiao, H.~Wang, Q.~Zhu, Y.~Ding, D.~J. Wild, Chem2bio2rdf:
  a semantic framework for linking and data mining chemogenomic and systems
  chemical biology data, BMC Bioinformatics 11~(1) (2010) 255.

\bibitem{celebi2019evaluation}
R.~Celebi, H.~Uyar, E.~Yasar, O.~Gumus, O.~Dikenelli, M.~Dumontier, Evaluation
  of knowledge graph embedding approaches for drug-drug interaction prediction
  in realistic settings, BMC bioinformatics 20~(1) (2019) 1--14.

\bibitem{hettige2019medgraph}
B.~Hettige, Y.-F. Li, W.~Wang, S.~Le, W.~Buntine, Medgraph: Structural and
  temporal representation learning of electronic medical records, in:
  Proceedings of the 24th European Conference on Artificial Intelligence, 2020.

\end{thebibliography}

\end{document}